\documentclass[prb,preprint,groupedaddress,showpacs]{revtex4}
\usepackage{graphicx}
\begin{document}
\bibliographystyle{prbrev}

\title{Evolution of the excited electron bubble in liquid
       $^4$He and the appearance of fission-like processes}

\author{David Mateo}
\affiliation{Departament ECM, Facultat de F\'{\i}sica,
and IN$^2$UB,
Universitat de Barcelona. Diagonal 647,
08028 Barcelona, Spain}

\author{Mart\'{\i} Pi}
\affiliation{Departament ECM, Facultat de F\'{\i}sica,
and IN$^2$UB,
Universitat de Barcelona. Diagonal 647,
08028 Barcelona, Spain}

\author{Manuel Barranco}
\affiliation{Departament ECM, Facultat de F\'{\i}sica,
and IN$^2$UB,
Universitat de Barcelona. Diagonal 647,
08028 Barcelona, Spain}

\begin{abstract}

We have studied the evolution of an excited 
electron bubble in superfluid $^4$He for several tens of picoseconds
combining the dynamics of the liquid with an adiabatic evolution for the electron.
The path followed by the excited bubble
in its decay to the ground state is shown to strongly depend on
pressure. While for pressures below 1 bar the 1P excited electron bubble
has allowance for radiatively decay to the deformed ground state,
evolving then non-radiatively towards the ground state of the spherical
electron bubble, we have found that above 1 bar two distinct baby
bubbles appear in the course of the dynamical evolution, pointing to a 
different relaxation path in which the electron may be localized in one 
of the baby bubbles while the other collapses, allowing for a pure 
radiationless de-excitation. Our calculations are in 
agreement with experiments indicating that relaxed 1P bubbles
are only observed for pressures smaller than a critical one, of the
order of 1 bar, and that above this value the decay of the excited
bubble has to proceed differently. A similar analysis carried out for 
the 2P bubble shows that the adiabatic approximation fails at an early 
stage of its dynamical evolution due to the crossing of the 2P and 1F
states.



\end{abstract}

\date{\today}

\maketitle

\section{Introduction}

Electron bubbles (ebubbles) in liquid helium are fascinating objects
with an apparently simple structure that have been the subject of
a large number of experimental and theoretical studies, see e.g. Refs.
\onlinecite{Ros06,Mar08,Nor67,Fow68,Miy70,Gri90,Gri92,Elo02,Gra06,Leh07,DuV69,Pi07}
and references therein.

The imaging of individual ebubbles moving in the liquid,\cite{Guo09} 
some unexplained events in cavitation experiments,\cite{Gho05} and the efforts 
in creating and detecting multielectron bubbles\cite{Tem07,Fan09} are
recent issues calling for a dynamical description of the electron
bubble, but not the only ones. For instance,
the equilibration of the electron bubble in superfluid liquid helium,
studied in detail by Eloranta and Apkarian\cite{Elo02} within
time-dependent density functional theory, also needed of
an accurate dynamical description.
The ebubbles addressed in that work are spherically symmetric, which
made the calculations affordable for using the best
available density functional (DF) for $^4$He, the so-called Orsay-Trento
(OT) functional.\cite{Dal95} Other dynamical studies have resorted to
much simpler approaches inspired on local functionals of the
kind proposed by Stringari and Treiner long time ago,\cite{Str87} or on
generalizations of the Gross-Pitaevskii equation to the
description of liquid helium.\cite{Ber00,Mad08,Jin10,Jin10b} 
They have allowed to carry out dynamical studies involving
non-spherical ebubbles, and their interaction with
vortices in the superfluid. However,
these local approaches do not describe the superfluid accurately.
In particular, its elemental excitations are poorly reproduced.
To circumvent this shortcoming, non-local extensions 
have been proposed\cite{Ber99} and  applied e.g., to vortex nucleation
in superfluid helium.\cite{Ber00b}

Another problem requiring a dynamical treatment,
still not addressed in full detail, is
the relaxation of an ebbuble after being excited by photoabsorption,
which constitutes the subject matter of this work.
This process couples the fairly slow displacement of the 
helium bubble with the rapid motion of the electron it hosts,
producing
excitations in the liquid that take away a sizeable part of the
energy deposited in the ebubble during the absorption.
The emission spectrum of the electron bubble
after it has  relaxed around the excited electron state
has been calculated.\cite{Mar03,Leh07,Mat10} However,
whether and how these full relaxed configurations are attained
before decaying by photoemission was not elucidated.

In this work we attempt a theoretical description of the evolution of
the excited ebubble based on the zero temperature DF approach, using an
as much accurate as technically feasible description of the liquid, and 
an electron-helium interaction that have been proved to reproduce the
experimental absorption energies of the ebubble. The initial
configuration is determined by a static
calculation of the excited ebubble. This state has a
large radiative lifetime, of the order of
several tens of microseconds, in contrast with
the short time scale for the helium displacement, of the order of
picoseconds. The subsequent dynamical evolution of the ebubble
is  described within the adiabatic approximation, which is
valid for a period of time difficult to ascertain,\cite{Els01} that we
shall discuss in some detail. We will show that, depending on the initial
excited state and the external pressure applied to the liquid, the
bubble may keep its initial simply connected topology, or evolve towards
a non-simply connected one made of two baby bubbles that share the
probability of finding the electron in,\cite{Jac01} the electron
eventually localizing in one of them while the
other collapses. To reduce the numerical effort to a reasonable
amount, we shall mostly discuss results for the collapse of an
ebubble starting from the spherical 1P state. Results
for the collapse of the 2P state will be also shown.

This work is organized as follows. In Sec. II we describe our
model and present a quasi-static study of the ebubble,
completing the results we have presented
elsewhere,\cite{Mat10} and recalling some technical details about the
method we have used to solve the variational equations for the fluid and
the electron. In Sec. III we present the adiabatic evolution
of the ebubble for two selected values of the liquid pressure.
The validity of the adiabatic approximation is analyzed in Sec. IV,
and a summary is presented in Sec. V.

\section{Quasi-static description}

We first address some properties of excited electron bubbles in
liquid $^4$He using the Orsay-Trento density functional
including the terms that mimic backflow effects 
and are crucial to quantitatively reproduce the experimental
phonon-roton dispersion relation in bulk liquid $^4$He at zero
pressure.\cite{Dal95}
They have no influence on the statics of the system,
and have been often neglected\cite{Anc03b,Pi07} in the dynamics
because their inclusion makes the dynamical calculations very
cumbersome.\cite{Elo02,Leh04,Gia03,Anc05} In practice,
we have found that these terms have little effect on the
dynamics of the electron bubble presented in this work.

The electron-helium (e-He) interaction has been modeled by the 
Hartree-type local effective potential derived by Cheng et
al.\cite{Che94} 
This allows us to write the energy of the electron-helium system  
as a functional of the electron wavefunction $\Phi({\bf r})$ and
the $^4$He effective macroscopic wavefunction
$\Psi({\bf r})= \sqrt{\rho({\bf r})} \exp [\imath S({\bf r})]$,
where $\rho({\bf r})$ is the particle density and
${\bf v}({\bf r}) = \hbar \nabla S({\bf r})/m_{He}$ is the velocity 
field of the superfluid:

\begin{eqnarray}
& & E[\Psi, \Phi] =
\frac{\hbar^2}{2\,m_{He}} \int d {\bf r}\, |\nabla \Psi({\bf r}\,)|^2 +
\int d {\bf r} \, \cal{E}(\rho)
\nonumber \\
&+ &\frac{\hbar^2}{2\,m_e} \int d {\bf r}\, |\nabla \Phi({\bf r}\,)|^2
+ \int d {\bf r}\, |\Phi|^2 V_{e-He}(\rho)  \; .
\label{eq1}
\end{eqnarray}
In this expression, ${\cal E}(\rho)$ is the $^4$He `potential' energy 
density, and the e-He interaction $V_{e-He}(\rho)$ is written as a 
function of the helium density.\cite{Che94} Details are given in Refs. 
\onlinecite{Gra06,Pi05}. In the absence of vortex lines, $S$ is zero 
and $E$ becomes a functional of $\rho$ and $\Phi$. Otherwise, one has 
to use the complex wavefunction $\Psi({\bf r})$ to describe the
superfluid.

For a given pressure ($P$), we have solved the
Euler-Lagrange equations which result from the variation
with respect to $\Psi^*$ and $\Phi^*$
of the zero temperature constrained grandpotential density
$\tilde{\omega}(\Psi, \Phi) = \omega(\Psi, \Phi)
- \varepsilon |\Phi|^2 $,  with
\begin{eqnarray}
\omega(\Psi, \Phi) &=&
\frac{\hbar^2}{2\,m_{He}} |\nabla \Psi({\bf r})|^2 +
{\cal E}(\rho)
\nonumber \\
&+ & \frac{\hbar^2}{2\,m_e} |\nabla \Phi|^2
+ |\Phi|^2 V_{e-He}(\rho) - \mu \rho \; ,
\label{eq2}
\end{eqnarray}
where $\mu$ is chemical potential of the liquid. The variation of 
the above functional yields two coupled equations that have to be 
selfconsistently solved
\begin{equation}
-\frac{\hbar^2}{2\, m_{He}}\Delta \Psi +
\left\{\frac{\delta \cal{E}}{\delta \rho} +|\Phi|^2\, \frac{\partial
V_{e-He}(\rho)}{\partial \rho} \right\} \Psi = \mu \Psi
\label{eq3}
\end{equation}
\begin{equation}
-\frac{\hbar^2}{2\, m_e}\Delta \Phi + V_{e-He}(\rho) \Phi  =
\varepsilon \Phi   \; ,
\label{eq4}
\end{equation}
where $\varepsilon$ is the eigenvalue of the Schr\"odinger
equation obeyed by the electron.

Our method of solving the variational equations is based 
on a high order discretization in Cartesian coordinates
of the differential operators
entering them (13-point formulas in the present case),
and the use of fast Fourier transformation 
techniques\cite{FFT} to efficiently compute the convolution integrals
in $\omega(\rho)$, such as the mean field helium potential
and the coarse-grained density entering the definition of
the correlation energy.\cite{Pi07}  This allows us to use
a large spatial mesh-step, of about 1 \AA {} size, without
an apparent loss of numerical accuracy when we compare our
results with others (see below) obtained using  3-point formulas
for the derivatives that, as a consequence, require a rather small
mesh-step to be accurate. The density at the boundary 
of the three-dimensional (3D) 140 \AA {} $\times$ 140 \AA {}
$\times$ 140 \AA {} box  used to carry out the calculations is 
fixed to the value of the bulk liquid density at the given $P$. 
We recall that knowledge of ${\cal E}(\rho)$ allows to determine the 
equation of state of the bulk liquid and its chemical potential, since 
$\mu = \partial {\cal E}/\partial \rho$ and 
$P = - {\cal E}(\rho) + \mu \, \rho$.
Eqs. (\ref{eq3}-\ref{eq4}) have been solved employing an imaginary
time method,\cite{Bar03} and we have carried out the appropriate tests
to check the stability of the solutions.
We mention that the energies we have obtained for the 
1S$\rightarrow$1P and 1S$\rightarrow$2P transitions\cite{Gra06} are in
very good agreement with experiment,\cite{Gri90} and that our results 
compare well with those obtained by Eloranta and Apkarian\cite{Elo02} 
using the same functional but a different numerical method and e-He 
interaction. This constitutes an excellent test not only for the numerics,
but also for the physical ingredients employed in both calculations.
We have recently discussed the effect of the presence of vortices 
on the absorption spectrum of ebubbles attached to them.\cite{Mat10}

Upon excitation to the 1P state by light absorption,
the ebubble experiences a drastic change of shape. This is due to the
fairly large radiative lifetime of this state (calculated to be 44
$\mu$s in Ref. \onlinecite{Mar03}, 60 $\mu$s in Ref. \onlinecite{Leh07},
and 56 $\mu$s in Ref. \onlinecite{Mat10}) as compared to any
characteristic helium timescale, allowing the liquid to relax around 
the excited state. As a consequence, the bubble adapts its shape 
to the 1P electron probability density before decaying by 
photoemission to the deformed 1S state. Consequently, the bubble
configuration at the emission time can be obtained by minimizing the
grandpotential of the system  keeping the electron in the excited 
1P state. We have done it by solving Eqs. (\ref{eq3}-\ref{eq4})
taking for $\Phi$ the $\Phi_{1{\rm P}}$ wavefunction. In this case,
a Gram-Schmidt scheme has been implemented to determine both the 1P
and 1S relaxed states that obviously no longer correspond to a spherical 
bubble. In this axially symmetric environment, the spherical nL 
states are split depending on the value ($m$) of the orbital angular
momentum on the symmetry $z$-axis, and the $\pm m$ states are 
degenerate. We have found that, within a nL manifold, the $m$ states 
are ordered in increasing $|m|$ values.\cite{Mat10} For this reason,
we will refer to the axially symmetric state that corresponds to
the $m=0$ submanifold when we speak of a deformed `nL' state.
When needed, we shall use the conventional notation for the orbital 
angular momentum of single particle states in linear molecules, 
namely $\sigma, \pi, \delta, \phi, \ldots$
for $|m| = 0, 1, 2, 3, \ldots$, and superscripts $+\,(-)$ for
specularly symmetric (antisymmetric) states.

\begin{figure}[f] 
\centerline{\includegraphics[width=8.5cm,clip]{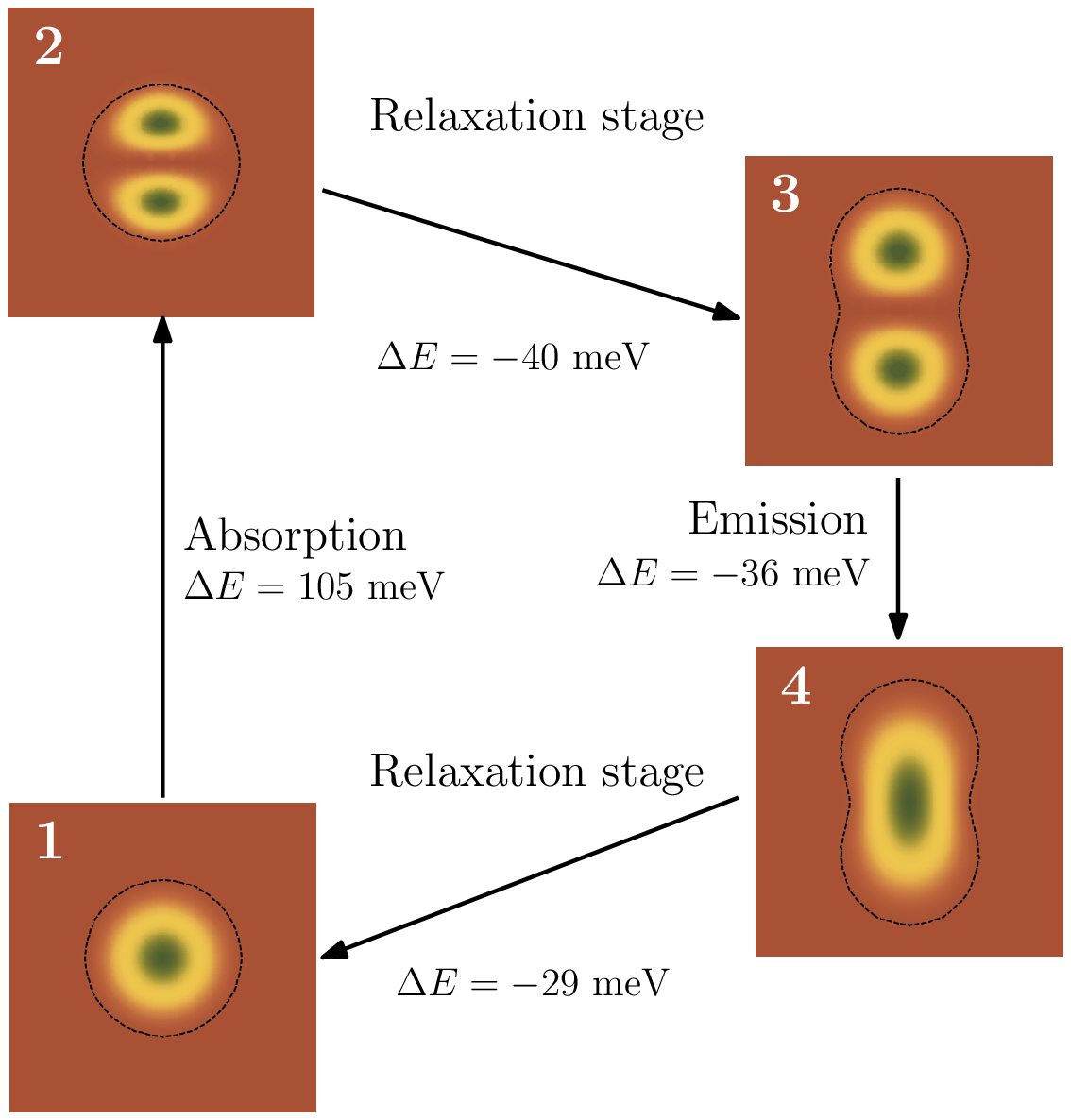}}
\caption{(Color online)
Ebubble quasi-equilibrium configurations
at different stages of the absorption-emission cycle 
corresponding to $P=0$.
The electron probability densities are represented by colored clouds.
The dashed line is the bubble dividing surface.
The size of the frames is  70 \AA{} $\times  70$ \AA{}.
}
\label{fig1}
\end{figure}

Figure \ref{fig1} displays quasi-equilibrium ebubble configurations
at different stages of the absorption-emission cycle obtained at $P=0$.
The electron probability densities are represented by colored clouds, these
with one lobe correspond to 1S states (spherical bubble, picture 1; 
deformed bubble, picture 4), and these with two lobes correspond to
1P states (spherical bubble, picture 2; deformed bubble, picture 3).
In this figure, 
the line indicates the bubble dividing 
surface, i.e., the surface at which the liquid density equals 
half the saturation density value $\rho_0$, e.g. 0.0218 \AA$^{-3}$
at $P=0$, and represents the surface of the helium bubble. 
We have found\cite{Mat10} that at $P=0$, the 1P$\rightarrow$1S emission
energy is 36 meV, close to the 35 meV found in Ref. \onlinecite{Leh07}, 
constituting another excellent test of the theoretical framework 
used by us and by these authors. The energy released in the optical
1S-1P absorption-emission cycle can be determined combining the results
we have obtained in previous works.\cite{Gra06,Mat10}
For instance, at $P=0$ the released energy is 69 meV
(compare with the 76 meV obtained in Ref. \onlinecite{Leh07}).
This energy is transferred to the superfluid through generation of
different kind of excitations.

\begin{figure}[f] 
\centerline{\includegraphics[width=8.5cm,clip]{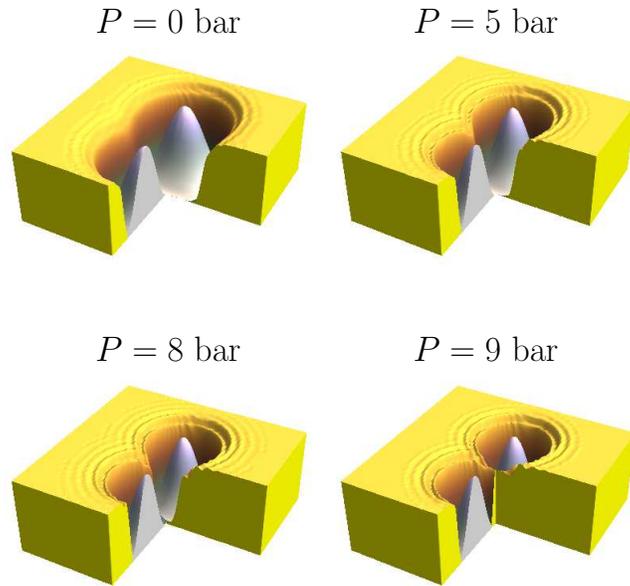}}
\caption{(Color online)
1P bubble quasi-equilibrium configurations for
$P=0$, 5, 8, and 9 bar.
Helium is represented by warm colors, and the electron
probability density (arbitrary scale) by cool colors.
The size of the samples is  70 \AA{} $\times  70$ \AA{}.
}
\label{fig2}
\end{figure}

Quasi-equilibrium configurations of the ebubble relaxed around
the 1P state are shown in Fig. \ref{fig2} for several $P$ values.
In this figure, helium is represented by warm colors, and the 
electron probability density (arbitrary units) by cool colors. The 
relaxation of the bubble around the 1P state produces a characteristic
two-lobe peanut structure whose waist -or neck- is progressively
marked as the pressure applied to the liquid increases. 
Notice that helium displays a stratified density around the bubble. 
This feature appears whenever the superfluid presents a kind of free 
surface, as in drops, films or bubbles.\cite{Dal95,Gra06,Bar03} 

Figure \ref{fig2} shows that
at a pressure of $\sim $ 8 bar helium starts to penetrate between the two
lobes of the electron wavefunction. At $\sim $ 9 bar the helium density in this region
reaches the saturation density, and the bubble splits into two
baby bubbles. This produces an abrupt change in the emission energy,
falling an order of magnitute between $P=8$ bar and $P=9$ bar; 
in the `broken neck' region extending up to the solidification
pressure, the photon emission energy is barely $\sim$ 1 meV.\cite{Mat10}
This is expectable at these pressures, as the main difference between 
the 1S and 1P probability densities appears in the waist region. If 
this region is inaccessible to the electron due to the presence of 
helium, these states become almost degenerate. On the contrary,
if this region is not accesible to the superfluid due, e.g., to
the presence of a vortex whose vorticity line coincides with the
symmetry axis of the ebubble, the baby bubbles may be held together 
by a tiny neck.\cite{Mat10}

It is worth pointing out that some of the quasi-equilibrium
configurations displayed in Fig. \ref{fig2} may not be reachable
in the evolution of the bubble.
The reason is that helium falling in the waist region during the 
violent collapse may produce a large pileup of superfluid in that
region, thus causing the actual breaking of the neck at
pressures much smaller than the 8 bar obtained quasi-statically.
This possibility has been anticipated by Maris.\cite{Mar08}

\section{Time evolution of the ebubble: picosecond dynamics}

The dynamics of the excess electron localization in liquid helium
has been adiabatically addressed by Rosenblit and Jortner using a
sharp surface model for the bubble.\cite{Ros95,Ros97}
The superfluid was considered as incompressible, and the bubble expansion time,
i.e., the time for creating the ebubble, was estimated to be 8.5 ps
when energy dissipation by emission of sound waves was taken into 
account.\cite{Ros97} This process exhibits a marked $P$ dependence, the higher 
the pressure the shorter the expansion time.

Later on, the incompressibility approximation was relaxed using a DF 
approach, finding that at $P=0$ the bubble surface breathes 
with a period of about 130 ps.\cite{Elo02} These calculations
have revealed that the localization process may launch shock waves, and that
the subsequent main dissipation mechanism is sound radiation; excitations in 
the roton well were not produced.\cite{Elo02} In the present work we consider that
the ebubble has had time enough to relax to its spherical 1S ground state and
the electron is subsequently excited by light absorption to the 1P state,
whose dynamical evolution is the subject matter of this section.

\subsection{Adiabatic time evolution}

Since the electron evolves much faster than helium as their mass
ratio is $m_{He}/m_e \simeq 7300$, we have followed the dynamics of the
excited ebubble by combining the actual time evolution of the liquid
with an adiabatic evolution for the electron. Within this approximation,
the electron wavefunction is found by solving, at each time step,
the static Schr\"odinger Eq. (\ref{eq4}) to obtain the instantaneous 1P
electron state $\Phi_{1{\rm P}}$, and the structure of the liquid is 
obtained by determining the complex, time-dependent effective
wavefunction $\Psi(\mathbf{r}, t)$ from the time-dependent DF equation
\begin{eqnarray}
& &\imath \hbar \,\frac{\partial \Psi(\mathbf{r},t)}{\partial t} =
-\frac{\hbar^2}{2\, m_{He}}\Delta \Psi(\mathbf{r}, t) 
\nonumber \\
&+&
\left\{ {\cal U}[\rho, \mathbf{v}] +|\Phi_{1{\rm P}}|^2\,
\frac{\partial V_{e-He}(\rho)}{\partial \rho} \right\} 
\Psi(\mathbf{r}, t) \;\; ,
\label{eq5}
\end{eqnarray}
where the effective potential $\cal{U}[\rho, \mathbf{v}]$ is given
e.g., in Refs. \onlinecite{Gia03,Leh04}
and has an explicit dependence on the local current field
$\mathbf{j}(\mathbf{r})=\rho(\mathbf{r})\,\mathbf{v}(\mathbf{r})$
arising from the backflow term the static potential energy in Eq.
(\ref{eq1}) lacks of.
These coupled equations are solved imposing as initial conditions the
stationary solution of the superfluid for the spherical 1S ebubble,
employing the electron wavefunction $\Phi_{1{\rm P}}$ for building
the e-He interacion, as indicated in Eq. (\ref{eq5}). The time step 
has been set to 0.01 ps, and we have used a fourth order
Runge-Kutta method to obtain the first time steps of the evolution.
To determine the solution for subsequent times, we have used
Hamming's (predictor-modifier-corrector) method.\cite{Ral60} This 
procedure is very robust and accurate, even for large amplitude 
motions.\cite{Bar06}

\subsection{Results}

We have solved the adiabatic-dynamic coupled equations for  $P=0$, 0.5, 1, 2, 3, 
and  5 bars. 
We will mostly show results for the two extreme pressure values, namely 0 and 5 bars.
The evolution starts by stretching the
bubble along the symmetry $z$-axis, and shrinking its waist.
This produces density waves in the liquid that take away a
sizeable part of the energy injected into the system during the
absorption process, 105 meV
at $P=0$ and 148 meV at 5 bar.\cite{Gra06}

The evolution can be safely followed for about 25-30 ps.
For larger times, the density waves reflected on the box
boundaries get back to the region where the bubble sits,
spoiling the calculation. This time interval is large enough to
see bubble splitting at the higher pressures.
Otherwise, one needs to introduce a source of damping in the
equation governing the liquid evolution [Eq. (\ref{eq5})],
to prevent sound waves from bouncing back.
Usually, introducing damping requires to enlarge the calculation box
to accomodate a buffer region where waves are washed out,
see e.g. Refs. \onlinecite{Elo02,Jin10,Jin10b,Cer85}.
This increases the number of grid points and slows the calculation.

Below 1 bar, we have found that the ebubble configuration is
simply connected and radiates a sizeable part of the excitation energy  
as sound waves. For instance, at $P=0$ bar, the energy difference
between the spherical 1P configuration and the relaxed 1P 
quasi-equilibrium configuration is $\sim 40$ meV (see Fig. \ref{fig1}).
The ebubble undergoes damped oscillations that will likely lead it to
the corresponding quasi-static 1P configuration described in
Sec. II. As a consequence, it would eventually decay radiatively 
to the deformed 1S state that will radiationless evolve towards the
spherical 1S state.

An example of this sort of evolution
is shown in Fig. \ref{fig3} for $P=0$.
We have found that after 15 ps, the shape of the 1P bubble is 
similar to the quasi-static configuration referred to in Sec. II.
Using a simpler model, Maris has found a smaller value,
11 ps.\cite{Mar08} The difference is
a natural consequence of the two basic approximations he has made, namely
treating the liquid as incompressible and neglecting sound wave radiation.
Due to the inertia of the bubble in the expansion process,
it continues to stretch in the direction of the symmetry axis. This dilatation
in the $z$-direction goes on for the largest times we have followed the evolution
(30 ps), accompanied by the appearance of a more marked neck.

At 1 bar, the neck
collapses due to the large kinetic energy of the liquid filling in the
region between the two 1P lobes, and the ebubble configuration becomes 
non-simply connected. This causes the -deformed- 1P and 1S levels to
become nearly degenerate, and their probability densities are almost
identical. The appearance of any asymmetric fluctuation, which is
beyond the scope and capabilities of our framework, will cause
the electron to eventually localize in either
of the baby bubbles. The subsequent evolution of the system is the collapse
of the empty baby bubble and the evolution of the other one towards the 
spherical 1S ground state. In this case, the excited 1P bubble decays 
to the 1S spherical configuration without passing through
the 1P quasi-static configuration described in the previous section, and
the de-excitation is non-radiative. An example of this sort of evolution
is shown in Fig. \ref{fig4} for $P=5$ bar. For this pressure, we have found
that a configuration similar to the simply connected quasi-static one is
attained after 10 ps, although the inertia of the bubble expansion 
breaks the quasi-static neck at about 18 ps. The density pileup in the
neck region continues and at about 22 ps the helium density in this region
has a peak of $\sim 2\rho_0$, whose relaxation pushes the two baby bubbles
in opposite directions helping the fission process.

Our calculations are in  agreement with cavitation
experiments\cite{Mar04} indicating that relaxed 1P bubbles
are only produced for pressures smaller than about 1 bar,
and that above this value the decay of the excited bubble 
has to proceed differently, likely radiationless. Indeed,
we have found that the 1P bubble fissions at $P=1$ bar, but it does not
at $P=0.5$ bar. Our results are
also in agreement with the interpretation\cite{Els01} of
the vanishing of the photoconductivity signal below 1 atm experimentally
observed by Grimes and Adams.\cite{Gri90} According to this
interpretation, an ebubble in the 1P state is unstable against a
radiationless de-excitation back to the ground state, the electron
ultimately settling into one of the baby bubbles while the other
collapses with phonon (`heat') emission. It is
this relased heat that drives the photocurrent.
Below that pressure, the ebubble decays
radiatively, it does not release enough heat, and is not detected in
the photocurrent experiment.

The evolution of the electron energies for the 1S (empty) and 1P
(occupied) states, together with a representation of the electron
probability densities, is presented in Fig. \ref{fig5} for two
pressure values.

The fission of the bubble at $P=1$ bar happens after 60 ps.
To obtain this result, we have proceeded as in Refs. 
\onlinecite{Jin10,Jin10b}, introducing a damping term in 
Eq. (\ref{eq5}). To make sure that the bubble does not fission
at $P=0.5$ bar, we have also introduced a damping term for 
this pressure.

We have studied the excitations produced in the liquid by
the expansion of the ebubble. 
From the evolution of the first wave front, we have estimated
that it moves in the $z$ direction at $\sim 330$ m/s at $P=0$, and at
$\sim 410$ m/s at $P=5$ bar.
These values are well above the speed of sound in helium at these
pressures, meaning that the dynamics is highly non-linear.
Besides, we have determined the nature of these excitations by
Fourier analyzing the density waves along the $z$ axis,
leaving aside the region near the bubble surface.
The density profile is shown in the top panel of Fig.
\ref{fig6}, corresponding to $P=0$ bar at 13 ps, and the
Fourier transform of the density fluctuation (related to the static
structure function of the liquid) is presented in the bottom panel.
Apart from the low-$q$ component, arising from the mean density profile,
one can identify two distinct peaks, the more marked one at $q \sim 0.8$
\AA$^{-1}$ in the phonon branch near to the maxon region, and
another at $q \sim 2.3$ \AA$^{-1}$ slightly to the right of the roton minimum.
A similar `roton' peak was found in Ref. \onlinecite{Elo02}. A less
marked peak appears at $q \sim 1.7$ \AA$^{-1}$,
slightly to the left of the roton minimum. Similar peaks have been found for
shorter and larger times. From the relative intensity of these peaks, we
are prone to identify most of the emitted waves as high energy
`phonons'.

We have also analyzed the effect of the backflow term on the 
appearance of the density waves. One can see from the bottom panel
of Fig. \ref{fig6}
that neglecting this term changes a little the relative intensity
of the phonon and roton peaks, increasing the former and decreasing the
later as expected from the effect of the backflow term
on the excitation modes of the superfluid, see Fig. 12 of Ref.
\onlinecite{Dal95}. We want to stress that rotons are not excited
if one uses a less accurate, local functional such
as that of Refs. \onlinecite{Str87,Ber00,Mad08,Jin10,Jin10b}. 
In this case, only the low-$q$
phonon spectrum of the superfluid is quantitatively reproduced. Whether
this has a sizeable influence on the ebubble dynamics or not, can only
be ascertained by a detailed comparison between the results obtained 
using both kind of functionals.

1P ebubbles may be excited by photoabsorption to the 1D state, either to
its $m=0$ component ($2\sigma^+$) or to its $m=\pm 1$ components ($1\pi^-$).
The absorption spectrum at different stages of the time evolution
can be measured in a pump-probe experiment by which
the ebubble is excited by two consecutive laser pulses. The delay set
between these pulses should correspond to
the time elapsed between the excitation of the spherical bubble and the
time at which the absorption spectrum of the 1P ebubble is recorded.
Time-resolved electronic spectroscopy has been proposed by
Rosenblit and Jortner as a tool for the exploration of the
localization dynamics of the excess electron.\cite{Ros97}

Time-resolved excitation energies are shown in Fig. \ref{fig7} at $P=0$
and 5 bar. While the evolution of the  $1\sigma^- \rightarrow 1\pi^-$ excitation 
is qualitatively similar at both pressures, the 
$1\sigma^- \rightarrow 2\sigma^+$ excitation evolves differently in the high 
pressure regime when the bubble splits.
Indeed, at zero bar the excitation energy smoothly decreases with time,
whereas at 5 bar it starts decreasing, increasing next, and eventually becoming larger than
the excitation energy to the $1\pi^-$ state. Note that both the change in behaviour and the 
crossing take place some picoseconds before bubble splitting.

The $1\sigma^- \rightarrow 1\pi^-$ transition is little affected by bubble splitting
because it involves two states with negative specular symmetry, which means that the probability density
of both states is zero in the neck region. 
On the contrary, the $1\sigma^- \rightarrow 2\sigma^+$
transition involves two states with different specular symmetry and thus it is more affected by 
bubble splitting.

Along with the excitation energies, some values of the associated oscillator 
strengths are displayed in Fig. \ref{fig7}. The oscillator strengths have been 
calculated in the dipole approximation as:\cite{Wei78}
$$
f_{ab} = \frac{2m_e}{3\hbar^2}(E_a-E_b)\big|\langle a | {\bf r} | b \rangle\big|^2
\; .$$
As known, the oscillator strengths fulfill a sum rule that in the one-electron 
case is $\sum_a f_{ab}=1$.\cite{note0}
At both pressures, we have found that these transitions have comparable
oscillator strengths. The largest difference appears for $P=5$ bar in the 
split-bubble regime. In it, the 
strength of the  $1\sigma^- \rightarrow 2\sigma^+$ transition is roughly half 
that of the  $1\sigma^- \rightarrow 1\pi^-$ transition.
We thus conclude that the analysis of the peak energy and oscillator strength
of the  $1\sigma^- \rightarrow 2\sigma^+$ transition might
disclose the fission-like behaviour of the excited 1P bubble, complementing
the experimental information gathered from cavitation and photoconductivity
experiments.

The current field [$\mathbf{j(r)}= \rho\mathbf{(r) v(r)}$]
is shown in Fig. \ref{fig8}
for $P=0$ and 5 bar at 12 and 22 picoseconds. At 12 ps
the current fields are qualitatively similar for both pressures: the bubble
expands along the symmetry axis and shrinks in a plane perpendicular to
it. At 22 ps, when $P=5$ bar,
large currents keep bringing liquid into the neck region, splitting the
bubble and producing important density oscillations in the central
region.

We have also followed the collapse of the 2P bubble at $P=0$.
About 2 ps after the collapse has started, a sizeable part of the
excitation energy has been released into the liquid and two waves 
are distinguishable around the bubble, as shown in  Fig. \ref{fig9}.
These waves travel through the liquid at the same speed as in the 
1P bubble case, $\sim 330$ m/s.
Shortly after 7 ps, the $m=0$ levels of the 2P and 1F states become very close
and, as discussed in the next section, see Eq. (\ref{eq7}), the adiabatic
approximation fails.
At this point, the 2P bubble displays an incipient four-lobe shape
arising from a similar structure in the 2P electron probability 
density. It is worth mentioning 
that a likely related effect, namely the near degeneracy
of the 2P and 1F states, was found in the quasi-static calculations of
Ref. \onlinecite{Mar03} as $P$ increased. 

In view of the mentioned failure and the lacking
of experimental information on the collapse of the 2P bubble, we have
closed its study at this point, leaving it for future work.

\section{Validity of the adiabatic approximation}

The validity of the adiabatic approximation
in the first stages of the bubble collapse, when the
topology of the bubble is simply connected, stems from
the very different time scale of the electron motion as
compared to that of the bubble,\cite{Mar00}
represented by the period of its shape oscillations. If
the fluid is incompressible and the bubble spherical,
the surface $\lambda$-modes of the cavity are at energies
\begin{equation}
\hbar \omega_{\lambda} =
\sqrt{\frac{\gamma}{m_{He} \rho_0 R^3}\,
(\lambda-1) (\lambda+1) (\lambda+2)} \;\; ,
\label{eq6}
\end{equation}
where $\gamma$ and $\rho_0$ are the surface tension and
atom density of the liquid, respectively.
For $\lambda=2$ this energy is about 1 K, and the period of
the oscillation is $ \sim 50$ ps.

The situation may change in the course of the
collapse because the energy difference 
$\Delta= E_{\rm 1P} -E_{\rm 1S}$
between the deformed states decreases and 
the time scale $\tau = h/\Delta$ may become similar to
the period of the shape oscillations of the deformed
bubble. Since $\Delta$ is small in the two bubble regime,
the approximation likely fails there.\cite{Els01,Rae01}
It is worthwhile mentioning that neck fluctuations,
not included in our approach nor in previous works,
would pinch off the bubble at earlier stages of the
collapse, in a similar way as they may cause the prompt scission
of the fissioning atomic nucleus after the saddle configuration
has been overcome.\cite{Bro89}

On the light of our model, in which no assumptions
are made on the shape of the bubble nor the 
impenetrability of the bubble surface by the localized electron,
it is instructive to analyze the validity of the
adiabatic approximation assuming that, during the evolution,
the bubble keeps its original axial symmetry
and specular symmetry about the plane perpendicular
to the symmetry axis that contains the node of the 1P state.
This excludes any possible fluctuation and the appearance of
asymmetric modes, like
the breathing mode discussed in Ref. \onlinecite{Els01}.
Our discussion relies on the detailed presentation by
Messiah,\cite{Mes62} that we summarize in the following.

The subsystem to which the adiabatic approximation is applied is the
electron, whose wavefunction is decoupled from that of the liquid.
This wavefunction evolves in the potential field generated by the 
liquid distribution, and its Hamiltonian is time-dependent,
${\mathcal H}_e(t) = {\mathcal H}_e[\rho_{He}(t)]$.
Let $\phi^t_n$ be an eigenfunction of the Hamiltonian at time $t$, so that
${\mathcal H}_e(t)\phi^t_n = \varepsilon_n(t)\phi^t_n$.
If $\phi_n(t)$ is the actual wavefunction $\phi^0_n$ evolved up to time $t$, one has
$ \phi_n(t) = {\mathcal U}(t)\phi^0_n $, where
${\mathcal U}$ is the evolution operator.
In the adiabatic approximation, one identifies $\phi_n(t)$ with
$\phi^t_n$, the intuitive justification being that if one perturbes 
the subsystem slowly and gently enough, it has enough time to adapt itself
to the new environment `with no inertia' from the past configuration.

The error made in this approximation for a given state $| i \rangle$
-the validity of the adiabatic approximation is assessed for
a given state of the subsystem, not necessarily for them all-
is defined as the probability of finding the subsystem in a
state different from  the initial one evolved in time within
the `true' dynamics, $\eta_{ij}=|\langle \phi^t_j|{\mathcal U}(t)|\phi^0_i\rangle |^2$.
This error can be written in a workable way as\cite{Mes62}
\begin{equation}
\eta_{ij}(t) = \left| \frac{\hbar}{[\varepsilon_j(t)-\varepsilon_i(t)]^2}
 \langle \phi^t_j |\frac{d{\mathcal H}_e}{dt} | \phi^t_i \rangle \right|^2 
\;\;  .
\label{eq7}
\end{equation}
If $\eta_{ij} \ll 1 \, \forall \, i \neq j$, the adiabatic approximation
is justified.
It is usually understood that it breaks
down when the levels get very close, or when they cross each other. 
Notice however that this assumes that these states can be
connected by the evolved Hamiltonian. If a symmetry is dynamically conserved
(in our case, angular momentum and specular symmetry are), then the adiabatic evolution
of states with a given quantum number associated to this symmetry is not
perturbed by states with different values of this quantum number.
Although sometimes ignored, this is a very reasonable statement.

In the case of the 1$\sigma^-$ state arising from the spherical 1P
manifold, the adiabatic approximation holds even when its energy becomes
almost identical to the energy of the 1$\sigma^+$ state 
arising from the spherical 1S one, i.e. a small $\Delta$ does not 
invalidate the adiabatic approximation.
The closest state having the same angular momentum and specular symmetry
is the 2$\sigma^-$ one arising from the spherical 1F manifold. 
At $P=5$ bar, we have found that these two states are 2000 K apart
in the 10-20 ps range. Since
$\hbar \sim$ 7.6 K\,ps, one has $\hbar/(E_{2\sigma^-}-E_{1\sigma^-})^2
\sim 2 \times 10^{-6}$ ps\,K$^{-1}$.  The value of the matrix element in
Eq. (\ref{eq7}) is some tens of kelvin per picosecond, so that 
the adiabatic approximation would be fulfilled even for the configuration
displayed in Fig. \ref{fig4} at 25 ps. Indeed, we
have calculated $\eta$ in the above time range and have found that it is
of the order of $10^{-8}$.

It is also worth analyzing the stability of the quasi-equilibrium configurations 
when the symmetries are not exactly conserved because of perturbations
from the environment. In this situation, let us assume that when the bubble splits
the electron localizes in one of the lobes. Leaving out the discussion on the actual
localization process, we have tried to infer the likely evolution of an ebubble
with a localized electron. The localized electron state in either baby bubble is
approximated by:
\begin{equation}
\Phi_{\pm} = \frac{1}{\sqrt{2}}\phi_{1{\rm S}} \pm \frac{1}{\sqrt{2}}\phi_{1{\rm P}}
\; \; .
\label{eq9}
\end{equation}
Consider now a short-time dynamics in which the liquid is kept frozen. 
The evolution of, e.g., the $\Phi_+$
localized state is an oscillation between the two lobes
\begin{equation}
\Phi(t) = e^{-iE_{1S} t / \hbar}[ \cos(\omega t)\Phi_+ -i \sin(\omega t)\Phi_- ]
\;\; , \label{eq10}
\end{equation}
where $\omega = (E_{1{\rm P}}-E_{1{\rm S}})/\hbar$. If this frequency is large enough,
the liquid cannot react to the localization of the electron in either lobe,
and will essentially behave as if the electron were delocalized.

The time elapsed between two consecutive localizations of the electron in
a given baby bubble is $\tau = \pi / \omega$. The value of this
period as a function of pressure for the quasi-equilibrium
configurations is displayed in Fig. \ref{fig10}.
In the split-bubble regime ($P\gtrsim 9$ bar), this period is of
several ps, indicating that the electron localization dynamics into
one of the baby bubbles is not a trivial process to address. The
electron will bounce back and forth as the liquid tries
to adapt to it.
Real time calculations are thus needed to describe electron
localization.

It is clear that the previous discussion on the validity of the
adiabatic evolution lacks for incorporating fluctuations or excitations
of low energy modes that may appear in the course of the bubble 
evolution and couple the 1P and 1S states that otherwise are not,
as previously discussed. One such mode has been thoroughly addressed by
Elser:\cite{Els01} a peanut configuration, whose walls are
impenetrable by the electron, is represented by two intersecting
sharp spheres of radius $R_2$ (instead of the deformed baby
bubbles displayed in Fig. \ref{fig4}) joined along a circular orifice
of radius $a$. These spheres are breathing in counterphase, producing
an antisymmetric mode whose stiffness ${\mathcal K}$ and inertia 
${\mathcal M}$ can be obtained analytically. This mode is very 
appealing, as it represents a small, swifting imbalance 
of the symmetric electron probability density.

In the harmonic limit, if $a \ll R_2$, the stiffness and inertia of the 
asymmetric mode are\cite{note2}
\begin{eqnarray}
{\mathcal K} &=& 16\left[ \frac{E^2_0}{R_2^2 \,\Delta} +
\pi (\gamma + R_2 P)
\right] 
\nonumber \\
{\mathcal M} &=& 4 \pi \xi\, m_{{\rm ^4He}} \,\rho_0 R_2^3  \;\; .
\label{eq11}
\end{eqnarray}
In these equations, $E_0=\gamma R_0^2 \sim 9$ meV represents the energy unit,
with $\gamma=2.36 \times 10^{-2}$ meV\AA$^{-2}$ being the surface tension 
of the liquid, $R_0 \sim$ 20 \AA{} is the radius of the spherical bubble,
$R_2 \sim$ 16 \AA{} is the radius of the baby bubbles,
$\xi \sim 1.70$ is a dimensionless constant, and 
$\Delta =E_{1{\rm P}} -E_{1{\rm S}}$. The frequency
of the antisymmetric breathing mode is given by $\omega_{AB}
= \sqrt{{\mathcal K}/{\mathcal M}}$, and the radius of the orifice is\cite{Els01}
\begin{equation}
\Delta
= \frac{4 \pi}{3}\, E_0 \left(\frac{R_0}{R_2}\right)^2 \left(\frac{a}{R_2}\right)^3
\; \; .
\label{eq12}
\end{equation}

The adiabatic approximation fails when $\omega_{AB} \simeq \Delta/\hbar$.
This yields $\Delta \sim 0.14$ meV in the $P=0-5$ bar range, as
only the first term in the stiffness turns out to be relevant
in this regime. Thus, keeping only the first term in ${\mathcal K}$,
one gets\cite{note3}
\begin{equation}
\frac{a}{R_2}
= \left[\left(\frac{27}{8 \pi^4 \xi}\right)
\,\left(\frac{m_e}{m_{He} \, \rho_0 R_0^3}\right)
\,\left(\frac{R_2}{R_0}\right)\right]^{1/9} \sim 0.13
\; \; .
\label{eq13}
\end{equation}
Hence, $a \sim 2.1$ \AA{}. Clearly, such analytical results cannot
be obtained within the DF approach, but we can use them to 
determine whether the dynamic and static configurations shown in
Secs. II and III are reliable.

\begin{figure}[f] 
\centerline{\includegraphics[width=8.5cm,clip]{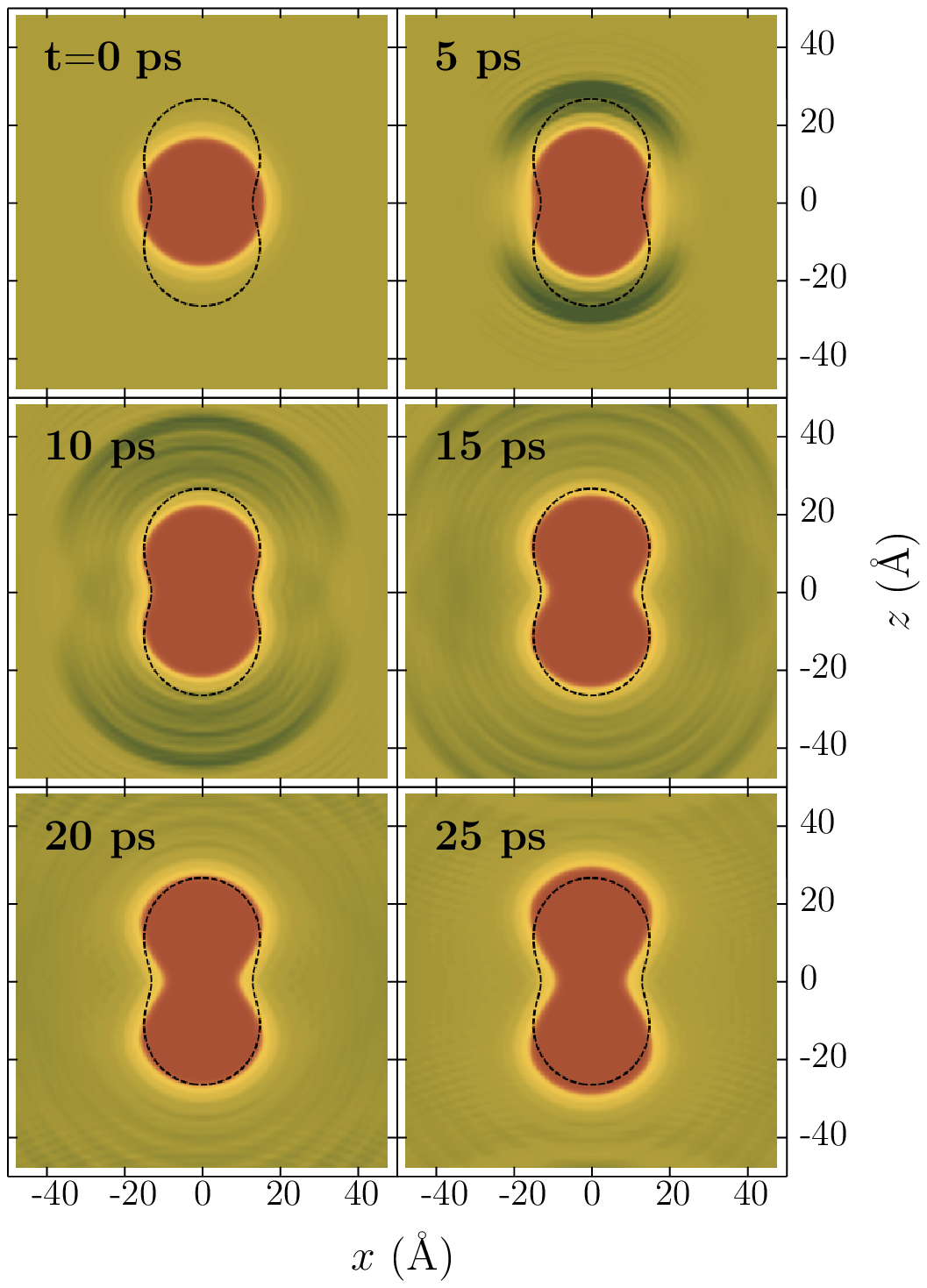}}
\caption{(Color online)
Adiabatic evolution of the 1P ebubble at $P=0$. The panels display
the helium configurations corresponding to 0, 5, 10, 15, 20, and 25 ps.
The dashed line represents the dividing surface of the quasi-equilibrium
configuration at $P=0$ shown in Fig. \ref{fig2}.
}
\label{fig3}
\end{figure}

\begin{figure}[f] 
\centerline{\includegraphics[width=8.5cm,clip]{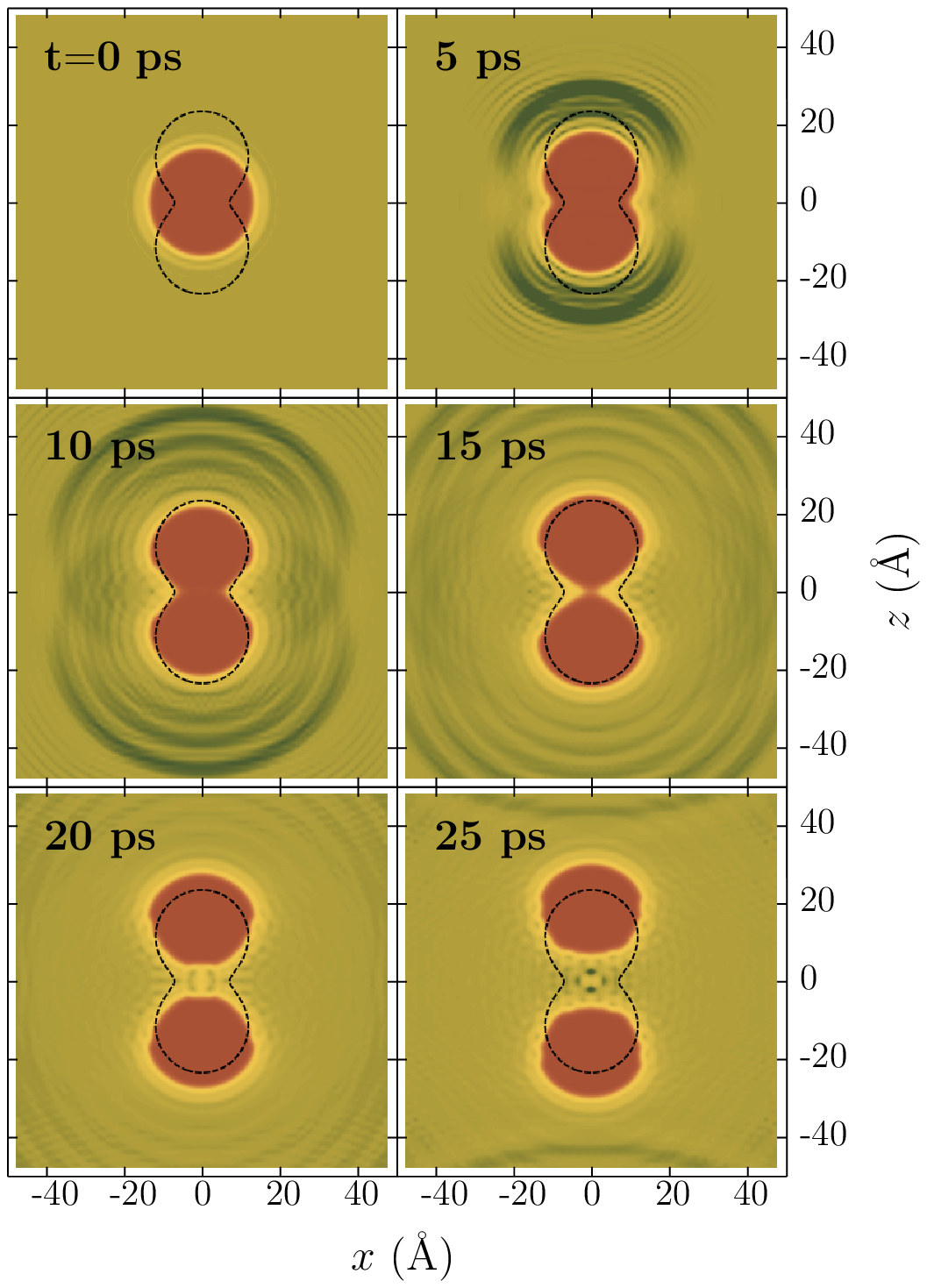}}
\caption{(Color online)
Adiabatic evolution of the 1P ebubble at $P=5$ bar. The panels display
the helium configurations corresponding to 0, 5, 10, 15, 20, and 25 ps.
The dashed line represents the dividing surface of the quasi-equilibrium
configuration at $P=5$ bar shown in Fig. \ref{fig2}.
}
\label{fig4}
\end{figure}

\begin{figure}[f] 
\centerline{\includegraphics[width=8.5cm,clip]{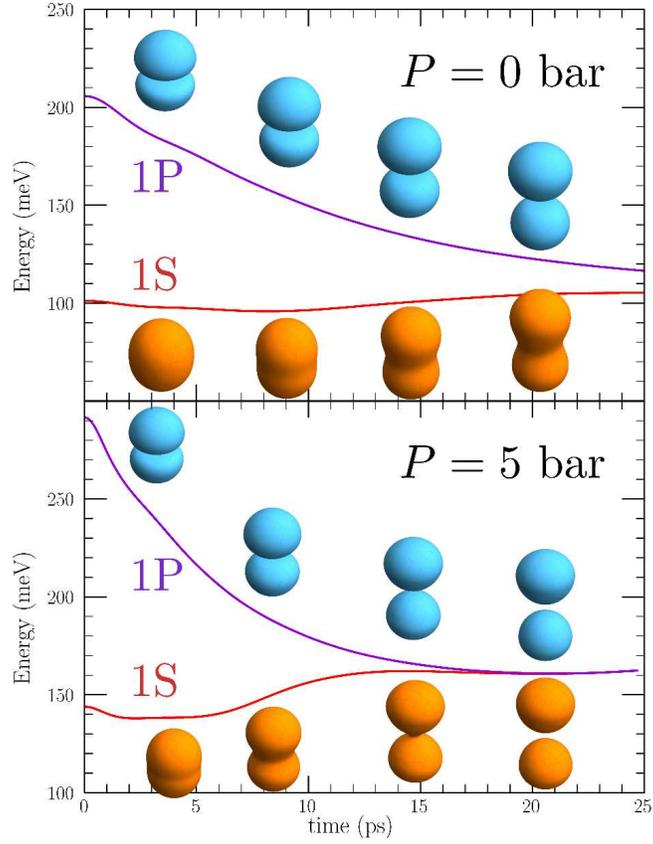}}
\caption{(Color online)
Adiabatic evolution of the energies of the 1S (empty) and 1P
(occupied) states, together with a representation of the electron
probability densities for $P=0$ (top panel) and $P=5$ bar (bottom panel).
The electron probability densities are also displayed at four
selected time values.
}
\label{fig5}
\end{figure}

\begin{figure}[f] 
\centerline{\includegraphics[width=8.5cm,clip]{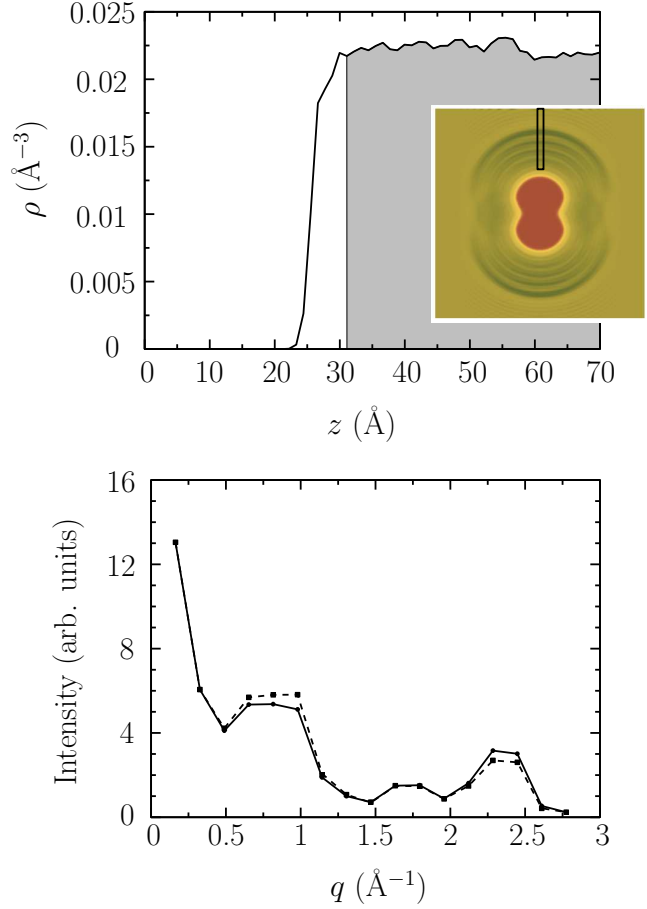}}
\caption{(Color online)
Top panel: Density profile of the superfluid corresponding to $P=0$ bar
at 13 ps (only the $z \geq 0$ part is shown).
The inset shows the ebubble configuration in a 140 \AA{} $\times$ 
140 \AA{} frame, and the region displayed in the rectangle is the
Fourier tranformed one.
Bottom panel: Solid line,
Fourier transform of the superfuid density profile shown in grey in 
the top panel. The dashed line is the Fourier transform of the density 
obtained at the same evolution time without including backflow effects.
In both cases
the peak at $q=0$ \AA$^{-1}$ arises from the mean density profile,
not subtracted from the local density before transforming.
The lines have been drawn as a guide to the eye.
}
\label{fig6}
\end{figure}

\begin{figure}[f] 
\centerline{\includegraphics[width=8.5cm,clip]{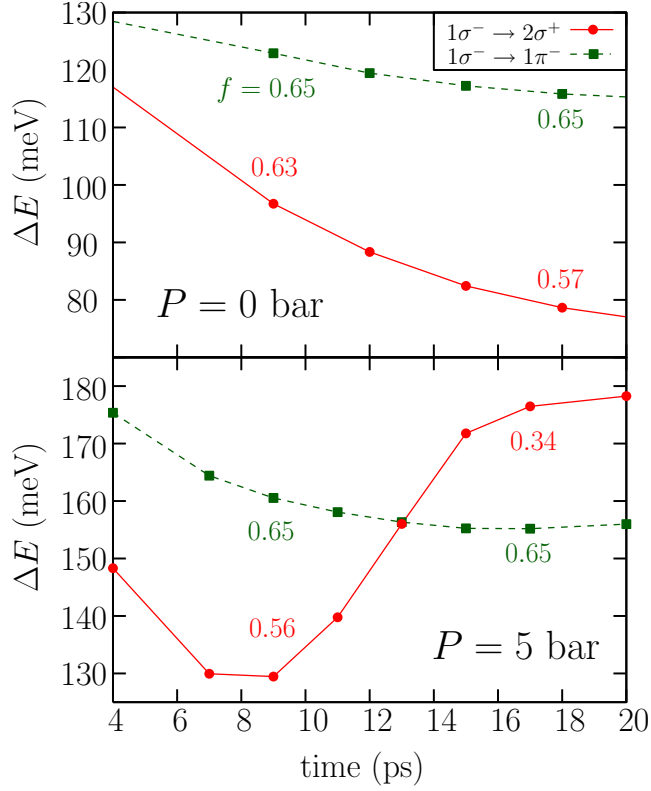}}
\caption{(Color online)
Time-resolved excitation energies at $P=0$ and $5$ bar
for electron transitions from the $1\sigma^-$ state arising from
the spherical 1P, $m=0$ one, to either the $1\pi^-$ state (circles),
or to the $2\sigma^+$ one (squares), both initially belonging to the
spherical 1D manifold. In each panel, the oscillator strength for two
selected time values is also displayed.
Lines have been drawn as a guide to the eye.
}
\label{fig7}
\end{figure}

\begin{figure}[f] 
\centerline{\includegraphics[width=10cm,clip]{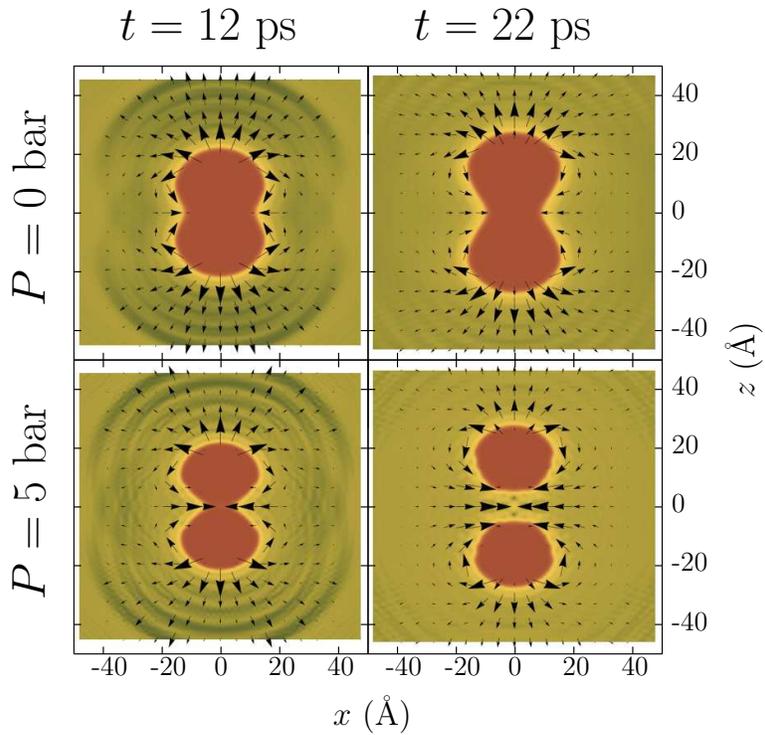}}
\caption{(Color online)
Current field
at $P=0$ (top panels) and 5 bar (bottom panels) at time $t=12$ ps (left)
and 22 ps (right). In each panel, the larger the arrow, the larger the
current.}
\label{fig8}
\end{figure}

\begin{figure}[f] 
\centerline{\includegraphics[width=8.5cm,clip]{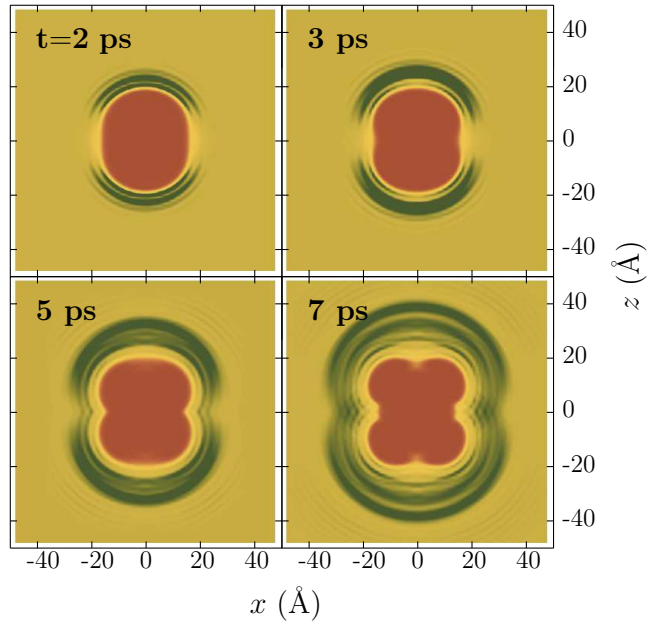}}
\caption{(Color online)
Adiabatic evolution of the 2P ebubble at $P=0$. The panels display
the helium configurations corresponding to 2, 3, 5, and 7 ps.
}
\label{fig9}
\end{figure}

\begin{figure}[f]
\centerline{\includegraphics[width=8.5cm,clip]{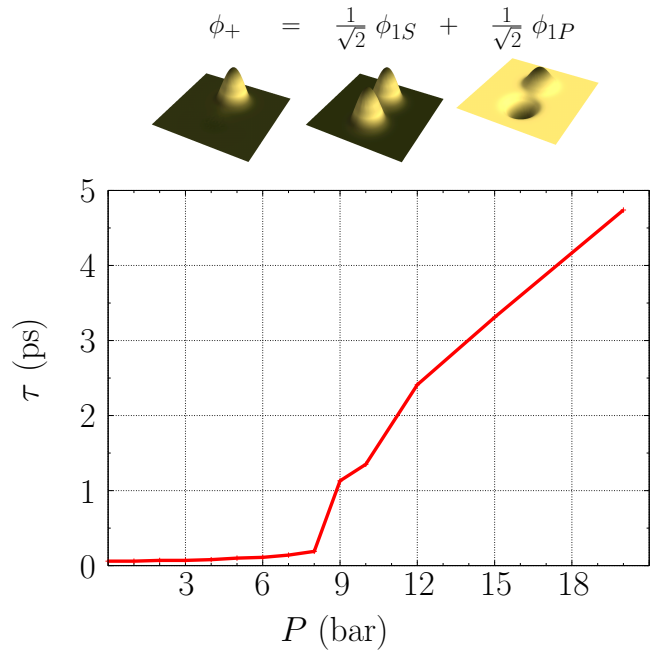}}
\caption{(Color online) Top:
superposition of the 1S and 1P states corresponding to the
quasi-equilibrium configuration at $P=9$ bar to  
approximately localize the electron in one of the baby bubbles.
Bottom:
tunneling period of a localized electron in one of the lobes
of the quasi-equilibrium ebubble as a function of pressure.
}
\label{fig10}
\end{figure}

The adiabatic approximation thus holds at
$P=0$, as the neck radius is fairly large, see Fig. \ref{fig3},
and $\Delta$ is always much larger than 0.14 meV, see Fig. \ref{fig5}.
From Fig. \ref{fig4} we also conclude that, at $P=5$ bar, the adiabatic
approximation is valid up to nearly the  collapse of the waist. 
Indeed, the neck radius of the helium configuration at 
about 17-18 ps is $\sim 2$ \AA{}, see Fig. \ref{fig4}.
 It is worth noting the difficulty in 
defining an effective radius for the orifice when the
surface of the bubble is diffuse; we recall that the surface
thickness of a $^4$He drop of 10$^3$-10$^4$ atoms is some 6-8
\AA{}.\cite{Har98} 
Note also that the surface thickness of the helium bubble is rather 
independent of the curvature of the surface, as can be inferred from the
fairly constant bright region around the bubbles displayed
in Figs. \ref{fig3} and \ref{fig4}.

Since we do not treat the bubble as impenetrable to the excess 
electron, the relation between the actual $\Delta$ and $a$
values should not exactly be as given by Eq. (\ref{eq12}).
Using the result $\hbar \omega_{AB} \leq \Delta= 0.14$ meV
as a criterion for the applicability
of the adiabatic approximation instead of reaching the limiting value $a
\sim 2$ \AA{}, we find that the approximation holds up to
21 ps, when the bubble has already split into two baby bubbles.
Both procedures indicate that when the adiabatic approximation likely
fails, the baby bubbles have already developped.

The previous analysis leads us to conclude that, at high pressures, baby
bubbles are formed some tens of picoseconds after the starting
of the collapse of the 1P bubble. From this point on,
the likely fate of the system is the localization of the 
electron in one of the baby bubbles and the collapse of the other.
This process is helped/triggered by fluctuations that break the
specular symmetry of the ebubble configuration. As mentioned,
determining the time scale of electron
localization is beyond the capabilities of the adiabatic approximation.
It has been calculated\cite{Jin10} that
once the electron is localized, it takes to the superfluid some 20 ps
to adapt to it while the other baby bubble is absorbed.

\section{Summary}

Within density functional theory, we have carried out an
analysis of the adiabatic evolution of the excited electron bubble
in superfluid liquid $^4$He. We have found that for pressures below
1 bar, the 1P ebubble may relax to its quasi-static equilibrium
configuration and eventually decay radiatively to the deformed 1S
state. This state evolves non-radiatively to the spherical 1S
bubble, completing the absorption/emission cycle. This conclusion 
arises in part from studies carried out for one hundred picoseconds
using a less accurate functional,\cite{JinUn} whose results 
qualitatively agree with ours for the first tens of picoseconds.

At higher pressures, the situation drastically changes and the
excited 1P bubble no longer decays to the quasi-static
equilibrium configuration, whose physical realization is unlikely. 
Indeed, our analysis of the adiabatic approximation indicates that it 
is valid up to a point where two deformed, nearly disconnected baby
bubbles appear in the dynamical evolution,
pointing towards a fission-like de-excitation process, the
likely subsequent evolution of the system being the  
localization of the electron in one of the baby bubbles 
and the collapse of the other. This collapse takes some 20 
ps,\cite{Jin10} and the whole de-excitation process is radiationless.

We have also found a marked change in the behavior of the time-resolved
absorption spectrum of the 1P bubble depending on whether the bubble
fissions or not, i.e., on the liquid pressure. This change
is in principle an experimentally accessible observable
whose determination may complement the information
obtained from cavitation and photoconductivity experiments.

Our analysis of the collapse of the 2P bubble  has shown
that the adiabatic approximation breaks down at an early stage of the
dynamical process due to the crossing of the 2P and
1F states. Although disclosed by the adiabatic approximation,
this crossing has nothing to do with the approximation itself, but
is inherent to the dynamics of the electron bubble. From the crossing
point on,
the bubble will relax around a mixed state with 2P and 1F components,
and hence the physical realization of a pure quasi-equilibrium 2P
configuration is unlikely. It is very plausible that the same applies 
to other high energy `nL' ebubbles generated in the absorption process.
The possibility that some of them undergo a spontaneous symmetry breaking, 
as suggested by Grinfeld and Kojima for the 2S state,\cite{Gri03}
can only reinforce our conclusion. Obviously, this does not question the
existence of either relaxed quasi-equilibrium configurations at low
pressures, or of baby bubbles at high pressures, arising from the 
evolution of the spherical 2P bubble. It just means that, on the one
hand, the relaxed bubble will not be a pure 2P configuration and, on the
other hand,  to study the de-excitation of these bubbles
one has to go beyond the adiabatic approximation and carry out a more 
demanding real time dynamics calculation for the electron.

\section*{Acknowledgments}

We thank  Humphrey Maris, Francesco Ancilotto, Dafei Jin,
and Alberto Hernando for useful discussions.
This work has been performed under Grants No. FIS2008-00421/FIS
from DGI, Spain (FEDER), and 2009SGR1289 from Generalitat de
Catalunya.

\pagebreak

\begin{thebibliography}{99}
\bibitem{note0}
Recall that if the initial state $| b \rangle$ is not the ground 
state, the sum includes negative terms arising from de-excitation 
transitions, hence a partial sum can be greater than unity.
\bibitem{note2}
Since the asymmetric mode of Ref. \onlinecite{Els01}
does not conserve the volume of the 
bubble, the stiffness slightly changes from the expression given in
Ref. \onlinecite{Els01} if $P \ne 0$. In the numerical
application we have used the $\gamma(P)$ law deduced in Ref.
\onlinecite{Gra06}.
\bibitem{note3}
This equation differs from Eq. (9) of Ref. \onlinecite{Els01}
because of an error made there to obtain it
from a straightforward combination of its Eqs. (3) and (8),
as can be easily checked.
\end{thebibliography}
\end{document}